\documentclass[10pt,letterpaper]{article}
\usepackage{opex3}
\usepackage{cite}

\begin{document}

\title{Coherent perfect absorption of path entangled single photons}
\author{Sumei Huang$^{1}$ and G. S. Agarwal$^{2,*}$}

\address{$^1$ Department of Electrical and Computer Engineering,\\
National University of Singapore, 4 Engineering Drive 3, 117583, Singapore\\$^2$ Department of Physics, Oklahoma State University, Stillwater, Oklahoma 74078, USA}
\email{*girish.agarwal@okstate.edu}


\begin{abstract}
We examine the question of coherent perfect absorption (CPA) of single photons, and more generally, of the quantum fields by a {\it macroscopic} medium. We show the CPA of path entangled single photons in a Fabry-Perot interferometer containing an absorptive medium. The frequency of perfect absorption can be
controlled by changing the interferometer parameters like the reflectivity and the complex dielectric constant of the
material. We exhibit similar results for path entangled photons in micro-ring resonators. For entangled fields like the
ones produced by a down converter the CPA aspect is evident in phase sensitive detection schemes such as in
measurements of the squeezing spectrum.
\end{abstract}



\section{Introduction}
\noindent The single photons have enabled us to understand many
aspects of the fundamentals of quantum physics \cite{Grangier1,Brien1,Aspect1,Aspect2,JPB}. For example
the availability of the single photon sources enabled the
detailed study of the Bohr's complementary principle and the
wave particle duality of light \cite{Brien2,Ostrowsky,Guo1}. Single photons are also considered to enable
different quantum information tasks \cite{Simon}. For example single photon
based quantum memories and single photon routers have been
demonstrated \cite{Zoller}. Clearly single
photons can be useful in many different applications. In this paper, we examine the possibility of  coherent perfect absorption (CPA) of single photons. The CPA was discovered by Wan et. al.~\cite{Cao,Stone} who pointed out that at certain frequencies which are determined by the intrinsic properties of the medium, the input light is completely absorbed. This effect arises from the interference of waves travelling to the right and left~\cite{SDG,Chong,Gmachl,Longhi,note,Noh}. Many systems have been shown to exhibit the CPA --- these include metamaterials~\cite{Feng,Klimov,Pu}, metal dielectric composites and metallic gratings~\cite{Agarwal,Gopal,Koh}, plasmonic nanostructures~\cite{Yoon,Lee,Ghenuche}. Many possible applications~\cite{Mock,Grote} of CPA in switches, modulators, and filters have been anticipated and hence a variety of systems have been examined for the existence of CPA. Clearly the CPA has aroused widespread interest. However all the studies so far have
used classical coherent fields. A natural question is--- is CPA possible with quantized fields.
Further does the existence of the CPA with quantized fields depend on the very nature of the
quantized fields. These are the questions we study in detail in this paper. Note that the single
photons themselves are incoherent so one might rule out the CPA with single photons. However, we can arrange single photons in a path entangled state to produce CPA of single photons.

The single photon can come from a
deterministic source or from a heralded source. A deterministic source is one where a single photon emitted on demand from single atoms \cite{Rempe}, quantum dots \cite{Kako,NC22}, and color centers in diamond \cite{Brien1}, whereas a heralded single photon is one of a pair of the entangled photons generated by the nonlinear process of spontaneously parametric down conversion in a nonlinear optical crystal \cite{JPB,Zavatta,NP,NC20}. The detection of one photon of a pair can herald the presence of the other one.
 We find that under the condition that the medium can exhibit CPA for classical light fields \cite{Cao,Stone,SDG,Chong,Gmachl,Longhi,Agarwal,Gopal,Koh}, the single photon can also be fully absorbed by the medium. This is an interference involving single photon which interferes with itself \cite{Dirac}. The incoming photon is in a path entangled state $(|1,0\rangle + \mathrm{e}^{i\phi}|0,1\rangle)/\sqrt{2}$ involving forward and backward paths. The situation with two or more photons is different as then the picture of interference changes \cite{Hong}. We present explicit results on
the perfect absorption of entangled photons and squeezed radiation fields.

The organization of the paper is as follows: In Sec. 2, we recall the main features of the CPA with classical coherent fields. In Sec. 3, we present the relation between the output quantum fields to the input quantum fields. The existence of the CPA depends on the nature of the correlations between the incoming quantum fields. In Sec. 4, we demonstrate the CPA of single photons in a path entangled state. In Sec. 5, we consider a general class of entangled fields and show the existence of the CPA in certain detection schemes, i.e., in measurements of physical quantities like squeezing. Our work considers absorption by a macroscopic medium and is distinct from the absorption by a microscopic entity like a single atom~\cite{Leuch,Pinotsi,Chuu,GYGuo}. The latter is much more challenging as the probability of absorption by a single atom is very low as compared to the absorption by a macroscopic medium.

\section{Coherent perfect absorption with classical fields}
\begin{figure}[htp]
\begin{center}
\scalebox{0.9}{\includegraphics{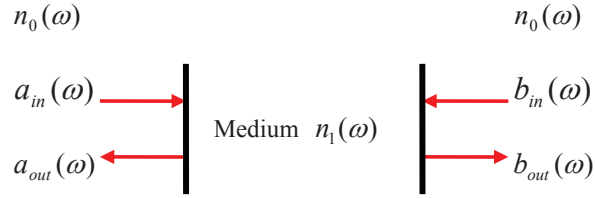}}
\caption{\label{Fig1} A Fabry-Perot interferometer. It is filled with an absorptive medium with complex refractive index $n_{1}(\omega)$. The refractive index outside the cavity is $n_{0}(\omega)$. The $a_{in}(\omega)$ and $b_{in}(\omega)$ are the incident counterpropagating quantum fields, and $a_{out}(\omega)$ and $b_{out}(\omega)$ are the output fields.}
\end{center}
\end{figure}
We consider a Fabry-Perot interferometer consisting of two partially transmitting mirrors separated by a distance $l$, as shown in Fig. \ref{Fig1}. A medium with complex refractive index $n_{1}(\omega)$ is placed inside the Fabry-Perot interferometer. The medium outside the cavity is the air with a refractive index $n_{0}(\omega)=1$. The two fields $a_{in}(\omega)$ and $b_{in}(\omega)$ are sent into the cavity from both sides of the cavity, respectively. If the two input fields are classical, the output fields $a_{out}(\omega)$ and $b_{out}(\omega)$ are given by
\begin{eqnarray}\label{1}
    a_{out}(\omega)&=&r_{L}(\omega)a_{in}(\omega)+t_{R}(\omega)b_{in}(\omega), \nonumber\\
    b_{out}(\omega)&=&r_{R}(\omega)b_{in}(\omega)+t_{L}(\omega)a_{in}(\omega),
\end{eqnarray}
where $r_{L}(\omega)$ and $r_{R}(\omega)$ are the reflection coefficients of $a_{in}(\omega)$ and $b_{in}(\omega)$, respectively, $t_{L}(\omega)$ and $t_{R}(\omega)$ are the transmission coefficients $a_{in}(\omega)$ and $b_{in}(\omega)$, respectively, the subscripts $L$ and $R$ stand for
the left and right sides of the interferometer, respectively. It has been shown that for a symmetric structure \cite{Dutta} $r_{L}(\omega)=r_{R}(\omega)=r\frac{1-e^{2ikn_{1}(\omega)l}}{1-r^2e^{2ikn_{1}(\omega)l}}$ and $t_{L}(\omega)=t_{R}(\omega)=\frac{(1-r^2)e^{ikn_{1}(\omega)l}}{1-r^2e^{2ikn_{1}(\omega)l}}$, where $k$ is the wavevector of light in vacuum with $k=2\pi/\lambda$ ($\lambda$ is the wavelength of the incident light), $r$ is the amplitude reflection coefficient of light for light incident from vacuum to $n_{1}(\omega)$ with $r=\frac{1-n_{1}(\omega)}{1+n_{1}(\omega)}$. Note that if the conditions
\begin{equation}\label{2}
    a_{in}(\omega) = b_{in}(\omega), \qquad r_{L}(\omega)+t_{R}(\omega)=0,
\end{equation}
are satisfied, the output fields disappear
\begin{equation}\label{3}
a_{out}(\omega)=b_{out}(\omega)=0.
\end{equation}
Thus the two incident counterpropagating coherent fields $a_{in}(\omega)$ and $b_{in}(\omega)$ are fully absorbed by the system. There is no output due to destructive interference between, say, the reflected part of $a_{in}$ and the transmitted part of $b_{in}$.

\section{Coherent perfect absorption of quantum fields}
Let us now consider the possibility of the coherent perfect absorption of the Bosonic quantum fields $\hat{a}_{in}(\omega)$ and $\hat{b}_{in}(\omega)$. These fields must satisfy the commutation relations
\begin{eqnarray}\label{4}
& &[\hat{a}_{in}(\omega), \hat{a}^\dag_{in}(-\Omega)]=2\pi\delta(\omega+\Omega),\nonumber\\
& &[\hat{b}_{in}(\omega), \hat{b}^\dag_{in}(-\Omega)]=2\pi\delta(\omega+\Omega),\nonumber\\
& &[\hat{a}_{in}(\omega), \hat{b}^\dag_{in}(-\Omega)]=0.
\end{eqnarray}
The Eqs.~(\ref{1}) cannot be used in the quantum theory as the medium inside the Fabry-Perot interferometer is absorptive. The absorption necessitates the use of quantum noise terms in (\ref{1})~\cite{Scully,GSAbook}. Thus we need to modify Eqs.~(\ref{1}) to
\begin{eqnarray}\label{5}
\hat{a}_{out}(\omega)&=&r_{L}(\omega)\hat{a}_{in}(\omega)+t_{R}(\omega)\hat{b}_{in}(\omega)+\hat{f}_{a}(\omega),\nonumber\\
\hat{b}_{out}(\omega)&=&r_{R}(\omega)\hat{b}_{in}(\omega)+t_{L}(\omega)\hat{a}_{in}(\omega)+\hat{f}_{b}(\omega),
\end{eqnarray}
where the additional noise terms $\hat{f}_{a}(\omega)$ and $\hat{f}_{b}(\omega)$ are introduced to make the output fields $\hat{a}_{out}(\omega)$ and $\hat{b}_{out}(\omega)$ satisfy the commutation relations $[\hat{a}_{out}(\omega),\hat{a}^\dag_{out}(-\Omega)]=2\pi \delta(\omega+\Omega)$, $[\hat{b}_{out}(\omega),\hat{b}^\dag_{out}(-\Omega)]=2\pi \delta(\omega+\Omega)$. They have the following commutation relations
\begin{eqnarray}\label{6}
& &[\hat{f}_{a}(\omega),\hat{f}_{a}^{\dag}(-\Omega)]=2\pi \{1-[|r_{L}(\omega)|^2+|t_{R}(\omega)|^2]\} \delta(\omega+\Omega), \nonumber \\
& &[\hat{f}_{b}(\omega),\hat{f}_{b}^{\dag}(-\Omega)]=2\pi\{1-[|r_{R}(\omega)|^2+|t_{L}(\omega)|^2]\} \delta(\omega+\Omega).
\end{eqnarray}
The mean values of  $\hat{f}_{a}(\omega)$ and $\hat{f}_{b}(\omega)$ are zeroes. And it is assumed that their normally ordered correlations are zeroes, $\langle \hat{f}_{a}^{\dag}(-\Omega)\hat{f}_{a}(\omega)\rangle=0$, $\langle \hat{f}_{b}^{\dag}(-\Omega)\hat{f}_{b}(\omega)\rangle=0$. Further, $\hat{f}_a$ and $\hat{f}_b$ cannot commute with each other. The requirement $[\hat{a}_{out}(\omega), \hat{b}^\dag_{out}(-\Omega)] = 0$ leads to
\begin{eqnarray}\label{7}
& &[\hat{f}_{a}(\omega),\hat{f}_{b}^{\dag}(-\Omega)] = 2\pi\delta(\omega+\Omega)\Gamma_{ab}(\omega), \nonumber \\
& &r_L(\omega)t^*_L(\omega) + t_R(\omega)r^*_R(\omega) + \Gamma_{ab}(\omega) = 0.
\end{eqnarray}
Thus the quantum noise terms in Eq.~(\ref{5}) are a must for the quantized theory although these do not contribute to the intensity measurements but can contribute to other measurements as discussed in Sec. 4.

We assume that the two input quantum fields have the following nonvanishing correlation functions
\begin{eqnarray}\label{8}
\langle \hat{a}^{\dag}_{in}(-\Omega)\hat{a}_{in}(\omega)\rangle &=& 2\pi N_{a}(\omega)\delta(\omega+\Omega),  \nonumber\\
\langle \hat{b}_{in}^{\dag}(-\Omega)\hat{b}_{in}(\omega)\rangle &=&2 \pi N_{b}(\omega)\delta(\omega+\Omega),
\end{eqnarray}
where the parameter $N_{u}(\omega)$ represents the number of photons at the frequency $\omega$ in the quantum field ($u=a,b$).
The power spectra of the output fields are defined as
\begin{eqnarray}\label{9}
\langle \hat{a}^{\dag}_{out}(-\Omega)\hat{a}_{out}(\omega)\rangle &=& 2\pi S_{aout}(\omega)\delta(\omega+\Omega),\nonumber\\
\langle \hat{b}^{\dag}_{out}(-\Omega)\hat{b}_{out}(\omega)\rangle &=& 2\pi S_{bout}(\omega)\delta(\omega+\Omega).
\end{eqnarray}
Using Eqs.~(\ref{5}) and (\ref{8}) and the correlation relation properties of $\hat{f}$, we find that the power spectra $S_{aout}(\omega)$ and $S_{bout}(\omega)$ of the output fields are given by
\begin{eqnarray}\label{10}
S_{aout}(\omega) &=& |r_{L}(\omega)|^2N_{a}(\omega)+|t_{R}(\omega)|^2N_{b}(\omega)\neq0,\nonumber\\
S_{bout}(\omega) &=& |r_{R}(\omega)|^2N_{b}(\omega)+|t_{L}(\omega)|^2N_{a}(\omega)\neq0.
\end{eqnarray}
Thus the input quantum fields $\hat{a}_{in}(\omega)$ and $\hat{b}_{in}(\omega)$ with no quantum correlations among them ($\langle \hat{a}_{in}^{\dag}(-\Omega)\hat{b}_{in}(\omega)\rangle=0$) cannot lead to perfect absorption. In the next section, we show how to modify the correlation functions (\ref{8}) so as to have perfect absorption of single photons. Thus if two photons are incident on the Fabry-Perot interferometer with one photon from each side then no CPA would occur.

\section{Coherent perfect absorption of path entangled photons}
\begin{figure}[htp]
\begin{center}
\scalebox{0.7}{\includegraphics{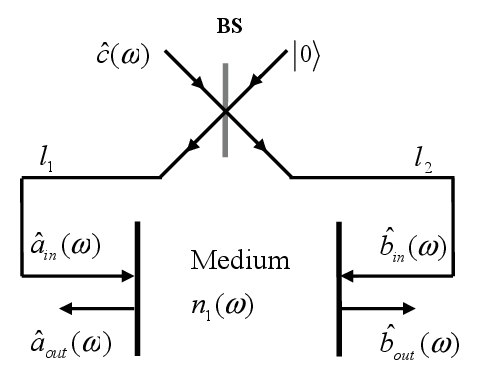}}
\caption{\label{Fig2}  A schematic for realizing the coherent perfect absorption of single photons in a Fabry-Perot interferometer. A quantum field $\hat{c}(\omega)$ mixes with a vacuum field $|0\rangle$ at a 50:50 beam splitter (BS). The output fields from the beam splitter respectively travel distances $l_{1}$ and $l_{2}$ to the double end mirrors of the Fabry-Perot cavity with a medium whose refractive index is $n_{1}(\omega)$.}
\end{center}
\end{figure}

\begin{figure}[htp]
\begin{center}
\scalebox{0.65}{\includegraphics{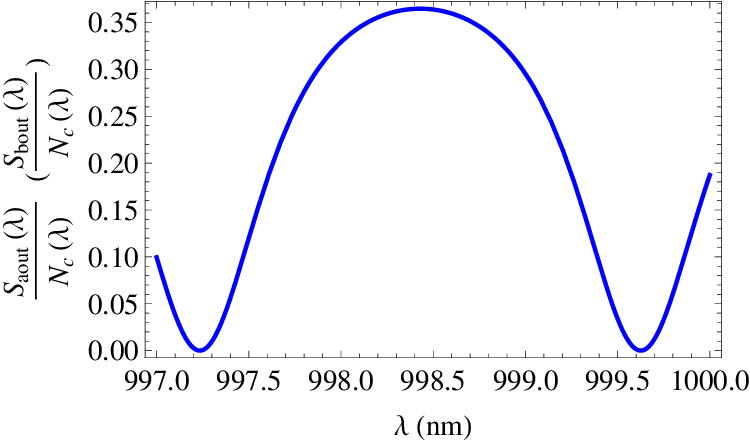}}
\caption{\label{Fig3}  The normalized power spectra $S_{aout}(\lambda)/N_{c}(\lambda)$ ($S_{bout}(\lambda)/N_{c}(\lambda)$) of the output fields as a function of wavelength $\lambda$, for the length of the cavity $l=115.79$ $\mu$m, the refractive index of the silicon medium is $n_{1}=3.6+0.0008i$ \cite{Cao}.}
\end{center}
\end{figure}

As previously mentioned, the key requirement to realize complete absorption of two classical lights incident on the two
opposite sides of the interferometer is the coherence between them. However, Dirac has pointed out that the single photons are not coherent and an interference is possible by a single photon interfering with itself \cite{Dirac}. Thus one would think of using an interferometric arrangement as shown in Fig. \ref{Fig2} to achieve perfect absorption of  single photons. A quantum field $\hat{c}(\omega)$ in a single photon Fock state is injected in the one input port of a 50:50 beam splitter, and a vacuum field $|0\rangle$ is injected in the other one. Then the output fields of the beam splitter travel through two distinct arms of different length $l_{1}$ and $l_{2}$, they become the quantum fields $\hat{a}_{in}(\omega)$ and $\hat{b}_{in}(\omega)$, entering the medium from the left and right sides. The photon entering the medium is in a path entangled state i.e., it is either in the forward path or in backward path. Such path entangled states are considered to be useful in teleportation \cite{Ralph}. If $\hat{d}(\omega)$ denotes the annihilation operator of the mode in the vacuum state $|0\rangle$,
the quantum fields $\hat{a}_{in}(\omega)$ and $\hat{b}_{in}(\omega)$ can be written in terms of the input operators $\hat{c}(\omega)$ and $\hat{d}(\omega)$ as
\begin{equation}\label{11}
\left(\begin{array}{c}
\hat{a}_{in}(\omega)\\
\hat{b}_{in}(\omega)
\end{array}\right)=\left(\begin{array}{cc}
e^{ikl_{1}}&0\\
0&e^{ikl_{2}}
\end{array}\right)\frac{1}{\sqrt{2}}\left(\begin{array}{cc}
1&i\\
i&1
\end{array}\right)\left(\begin{array}{c}
\hat{c}(\omega)\\
\hat{d}(\omega)
\end{array}\right),
\end{equation}
where $\hat{c}(\omega)$ and $\hat{d}(\omega)$ have the following correlation functions
\begin{eqnarray}\label{12}
\langle \hat{c}^{\dag}(-\Omega)\hat{c}(\omega)\rangle&=&2\pi N_{c}(\omega)\delta(\omega+\Omega),\nonumber\\
\langle \hat{d}^{\dag}(-\Omega)\hat{d}(\omega)\rangle&=&0,
\end{eqnarray}
where $N_{c}(\omega)$ is the photon number in the quantum field $\hat{c}(\omega)$. From Eq. (\ref{11}), one gets
\begin{eqnarray}\label{13}
\hat{a}_{in}(\omega)&=&\frac{1}{\sqrt{2}}[\hat{c}(\omega)+i\hat{d}(\omega)]e^{ikl_{1}},\nonumber\\
\hat{b}_{in}(\omega)&=&\frac{1}{\sqrt{2}}[i\hat{c}(\omega)+\hat{d}(\omega)]e^{ikl_{2}},
\end{eqnarray}
thus the quantum fields  $\hat{a}_{in}(\omega)$ and $\hat{b}_{in}(\omega)$ are path entangled.
Furthermore, by adjusting the lengths of the two arms so that the length difference $l_{1}-l_{2}$ satisfies $e^{ik(l_{1}-l_{2})}=i$, the correlation functions for $\hat{a}_{in}(\omega)$ and $\hat{b}_{in}(\omega)$ become
\begin{eqnarray}\label{14}
\langle \hat{a}_{i}^{\dag}(-\Omega)\hat{a}_{j}(\omega)\rangle&=&2\pi \frac{N_{c}(\omega)}{2}\delta(\omega+\Omega),
\end{eqnarray}
where $\hat{a}_{i}$ can be either $\hat{a}_{in}$ or $\hat{b}_{in}$. It is worth noting that the normally ordered auto-correlation functions and the normally ordered cross-correlation functions of the fields $\hat{a}$ and $\hat{b}$  are identical.
In this case, the power spectra of the output fields $S_{aout}(\omega)$ and $S_{bout}(\omega)$ defined in Eq. (\ref{9}) are found to be
\begin{eqnarray}\label{15}
S_{aout}(\omega)&=&\frac{N_{c}(\omega)}{2}|r_{L}(\omega)+t_{R}(\omega)|^2,\nonumber\\
S_{bout}(\omega)&=&\frac{N_{c}(\omega)}{2}|r_{R}(\omega)+t_{L}(\omega)|^2.
\end{eqnarray}
These are to be compared with the result (\ref{10}). The result (\ref{15}) is obtained as a photon interferes with itself.
Since $r_{R}(\omega)=r_{L}(\omega)$ and $t_{R}(\omega)=t_{L}(\omega)$, one has $S_{aout}(\omega)=S_{bout}(\omega)$.
When the condition
\begin{eqnarray}\label{16}
r_{L}(\omega)+t_{R}(\omega)=0,
\end{eqnarray}
is satisfied, the power spectra of the output fields $S_{aout}(\omega)$ and $S_{bout}(\omega)$ vanish
\begin{eqnarray}\label{17}
S_{aout}(\omega)=S_{bout}(\omega)=0.
\end{eqnarray}
Therefore, the path entangled photons lead to no output due to the quantum destructive interference between them. We have thus shown the occurrence of the CPA of single photons. For the case of the Fabry-Perot cavity containing Si~\cite{Cao}, we show the normalized power spectra $S_{aout}(\lambda)/N_{c}(\lambda)$ ($S_{bout}(\lambda)/N_{c}(\lambda)$) as a function of wavelength $\lambda$ in Fig. \ref{Fig3}. One can see that $S_{aout}(\lambda)/N_{c}(\lambda)=S_{bout}(\lambda)/N_{c}(\lambda)=0$ at $\lambda=997.234$ nm, $999.625$ nm, which is consistent with Eq. (\ref{17}). We note that one can choose a variety of materials to produce CPA with single photons at different wavelengths.
For example, the metal dielectric composites can produce CPA in the visible \cite{Agarwal,Gopal,Koh}. Note that the CPA occurs for specific frequencies---hence if the incident radiation is broadband then the transmitted radiation will have holes at the frequencies where CPA occurs. We note that although the figure \ref{Fig3} is like the figure for input coherent fields, the single photon nature can be verified by measurements of the intensity-intensity correlation
function $G_2$ \cite{Grangier1}. The CPA is achieved by using path entangled photons otherwise as shown in the previous section no CPA will occur for single photons.

\begin{figure}[htp]
\begin{center}
\scalebox{0.8}{\includegraphics{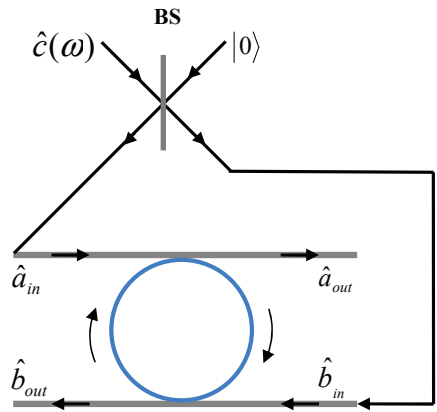}}
\caption{\label{Fig4}  The coherent perfect absorption of path entangled photons in a micro-ring resonator coupled to two waveguides.}
\end{center}
\end{figure}

As another example of perfect absorption of path entangled photons, let us consider a micro-ring resonator with two input-output waveguide couplers~\cite{Zhu} as shown in Fig.~\ref{Fig4}. We select a single mode of the resonator. Such systems are being proposed for optical switching and modulator applications~\cite{Mock,Grote}. The quantum theory of the fields in resonators is described in text books~\cite{Wallsbook0}. In what follows we generalize it by including input fields from the two sides and the internal loss.
The quantum field $\hat{a}(t)$ in the optical resonator with frequency $\omega_c$ satisfies
\begin{equation}\label{18}
  \dot{\hat{a}} = -(i\omega_c+\kappa_i+2\kappa)\hat{a} + \sqrt{2\kappa}\hat{a}_{in} + \sqrt{2\kappa}\hat{b}_{in}+\sqrt{2\kappa_i}\hat{f},
\end{equation}
where $\kappa_i$ is the internal loss rate in the micro-ring resonator, $\kappa$ is the reactive coupling rate between each waveguide
and the resonator, $\hat{a}_{in}$ and $\hat{b}_{in}$ are the input quantum fields, and $\hat{f}$ is the input vacuum noise associated with the internal loss rate $\kappa_i$, it has zero mean value and nonvanishing correlation function
\begin{equation}\label{19}
\langle \hat{f}(t)\hat{f}^{\dag}(t')\rangle=\delta(t-t').
\end{equation}
The Fourier transform of Eq.~(\ref{18}) yields
\begin{equation}\label{20}
  \hat{a}(\omega) = \frac{\sqrt{2\kappa}[\hat{a}_{in}(\omega) + \hat{b}_{in}(\omega)]+\sqrt{2\kappa_i}\hat{f}(\omega)}{2\kappa+\kappa_i+i(\omega_c-\omega)}.
\end{equation}
The fields at the output ports are given by~\cite{Wallsbook}
\begin{eqnarray}\label{21}
& &\hat{a}_{out}(\omega) = -\sqrt{2\kappa}\hat{a}(\omega) + \hat{a}_{in}(\omega), \nonumber\\
& &\hat{b}_{out}(\omega) = -\sqrt{2\kappa}\hat{a}(\omega) + \hat{b}_{in}(\omega).
\end{eqnarray}
Using the quantum correlations given by Eq.~(\ref{14}) and Eq.~(\ref{19}), and Eqs.~(\ref{20})--(\ref{21}), we find that the power spectra of the output fields $S_{aout}(\omega)$ and $S_{bout}(\omega)$ are
\begin{equation}\label{22in}
  S_{aout}(\omega) = S_{bout}(\omega) = \frac{N_{c}(\omega)}{2}\frac{(\kappa_{i}-2\kappa)^2+(\omega-\omega_{c})^2}{(\kappa_{i}+2\kappa)^2+(\omega-\omega_{c})^2}.
\end{equation}
When $\omega=\omega_{c}$, one has
\begin{equation}\label{22}
  S_{aout}(\omega_c) = S_{bout}(\omega_c) = \frac{N_{c}(\omega_{c})}{2}\frac{(\kappa_{i}-2\kappa)^2}{(\kappa_{i}+2\kappa)^2}.
\end{equation}
Note that $S_{aout}(\omega_c) = S_{bout}(\omega_c)=0$ if $\kappa_i=2\kappa$. The latter $(\kappa_i=2\kappa)$ is the critical coupling condition~\cite{SDG,Chong,Gmachl,Longhi,Vahala}. We thus demonstrate that a ring resonator under critical coupling condition leads to the CPA of single photons of frequency equal to the frequency of the resonator mode. The critical coupling is important as in other studies \cite{SDG,Chong,Gmachl,Longhi,note,Noh}.

\section{Homodyne Detection and Coherent Perfect Absorption of Entangled Fields}
So far we considered the question of CPA of single photons. We based our analysis on the interference. For fields more than one photon, a variety of interference effects are possible \cite{GSAbook}. For two single photons one coming from each side we would have Hong-Ou-Mandel effect \cite{Hong} whereby the probability of seeing one photon at each output can go to near zero, however one would still have nonzero probability of seeing two photons on either side \cite{ft,SDGOL}. Clearly the observed behavior depends on what detection scheme we choose. For entangled fields as for example those produced by the parametric down converter, the quantum mechanical interference effects are best studied by homodyne detection \cite{Agarwalprl,Xiao,Zhang}. In particular the squeezing spectrum~\cite{Wallsbook0} is a very good signature of the quantum interference. We show that the squeezing spectrum for the output fields exhibits a well defined minimum which can go to below zero at the point where perfect single photon absorption occurs.

\begin{figure}[htp]
\begin{center}
\scalebox{0.9}{\includegraphics{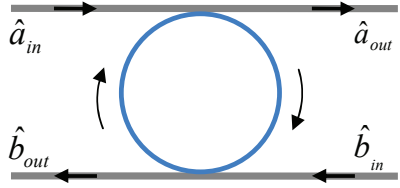}}
\caption{\label{Fig5}  The coherent perfect absorption of path entangled photons in a micro-ring resonator coupled to two waveguides.}
\end{center}
\end{figure}

 We consider the configuration of the resonator with different input fields as shown in Fig.~\ref{Fig5}. The $\hat{a}_{in}(\omega)$ and $\hat{b}_{in}(\omega)$ operators in Fig.~\ref{Fig5} describe the two modes of the squeezed vacuum field. Let us assume that the squeezed vacuum field has a central frequency $\omega_l$. Then writing $\hat{a}_j(t) = \hat{\tilde{a}}_j(t)\mathrm{e}^{-i\omega_l t}$, the slowly varying part $\hat{\tilde{a}}_j$ has the following nonvanishing correlation functions~\cite{Gardiner},
\begin{eqnarray}\label{23}
    & &\langle \hat{\tilde{a}}_{in}(\omega)\hat{\tilde{b}}_{in}(\Omega)\rangle = \langle \hat{\tilde{b}}_{in}(\omega)\hat{\tilde{a}}_{in}(\Omega)\rangle= 2\pi M\frac{\Gamma^2}{\Gamma^2+\omega^2}\delta(\omega+\Omega)\mathrm{e}^{i\varphi}, \nonumber\\
    & &\langle \hat{\tilde{a}}_{in}(\omega)\hat{\tilde{a}}_{in}^{\dag}(-\Omega)\rangle = \langle \hat{\tilde{b}}_{in}(\omega)\hat{\tilde{b}}_{in}^{\dag}(-\Omega)\rangle= 2\pi(N\frac{\Gamma^2}{\Gamma^2+\omega^2}+1)\delta(\omega+\Omega),
\end{eqnarray}
where $N=\sinh^2 r_{s} $, $M=\sinh r_{s} \cosh r_{s} $, $r_{s}$ is the squeezing parameter and $\varphi$ is the phase of the squeezed vacuum field, and $\Gamma$ is the linewidth of the squeezed vacuum field. It is noted that the two-photon cross-correlation functions between the quantum fields $\hat{\tilde{a}}_{in}(\omega)$ and $\hat{\tilde{b}}_{in}(\omega)$ are nonzero, however, correlations like $\langle \hat{\tilde{a}}_{in}^\dag(-\Omega) \hat{\tilde{b}}_{in}(\omega)\rangle$ are zero. We, therefore, calculate the spectra of the $\hat{x}$ quadratures of the output fields here, which are defined as
\begin{equation}\label{24}
\langle \hat{\tilde{x}}_{uout}(\omega) \hat{\tilde{x}}_{uout}(\Omega)\rangle=2\pi S_{uout}^{x}(\omega)\delta(\omega+\Omega),
\end{equation}
where $\hat{\tilde{x}}_{uout}(\omega)=\frac{1}{\sqrt{2}}[\hat{\tilde{u}}_{out}(\omega) + \hat{\tilde{u}}_{out}^{\dag}(-\omega)]$, $(u=a,b)$.

\begin{figure}[htp]
\begin{center}
\scalebox{0.65}{\includegraphics{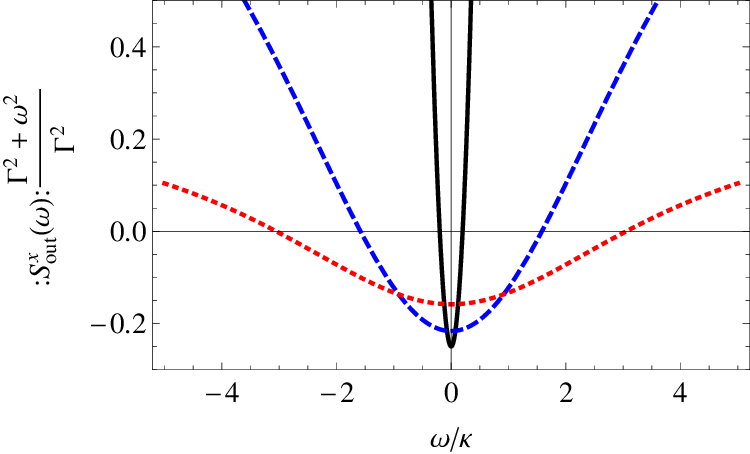}}
\caption{\label{Fig6} The normalized spectrum $:S_{out}^{x}(\omega):\times\frac{\Gamma^2+\omega^2}{\Gamma^2}$ of the $\hat{x}$ quadratures of the output fields as a function of the normalized frequency $\omega/\kappa$ for different values of squeezing parameters $r_{s}=0.5$ (dotted), 1 (dashed), 3 (solid). Parameter: $\varphi=0$.}
\end{center}
\end{figure}

For the optical mode $\hat{a}$ and the noise operator $\hat{f}$ in Eq.~(\ref{18}), we also introduce the slowly moving operators with tildes, $\hat{a}(t)=\hat{\tilde{a}}(t)\mathrm{e}^{-i\omega_l t}$, $\hat{f}(t)=\hat{\tilde{f}}(t)\mathrm{e}^{-i\omega_l t}$. Moreover, we assume the central frequency $\omega_{l}$ of the squeezed vacuum field is equal to the optical resonance frequency $\omega_{c}$. Then the quantum field $\hat{\tilde{a}}(\omega)$ in the resonator is given by
\begin{equation}\label{25}
\hat{\tilde{a}}(\omega)=\frac{\sqrt{2\kappa}[\hat{\tilde{a}}_{in}(\omega)+\hat{\tilde{b}}_{in}(\omega)]+\sqrt{2\kappa_i}\hat{\tilde{f}}(\omega)}{\kappa_{i}+2\kappa-i\omega},
\end{equation}
where $\hat{\tilde{a}}(\omega)$ and $\hat{\tilde{f}}(\omega)$ are the Fourier transform of $\hat{\tilde{a}}(t)$ and $\hat{\tilde{f}}(t)$, respectively. In this case,
the output fields in Eq.~(\ref{21}) become
\begin{eqnarray}\label{26}
\hat{\tilde{a}}_{out}(\omega)&=&\frac{-2\kappa\hat{\tilde{b}}_{in}(\omega)+(\kappa_{i}-i\omega)\hat{\tilde{a}}_{in}(\omega)-2\sqrt{\kappa\kappa_{i}}\hat{\tilde{f}}(\omega)}{\kappa_{i}+2\kappa-i\omega},\nonumber\\
\hat{\tilde{b}}_{out}(\omega)&=&\frac{-2\kappa\hat{\tilde{a}}_{in}(\omega)+(\kappa_{i}-i\omega)\hat{\tilde{b}}_{in}(\omega)-2\sqrt{\kappa\kappa_{i}}\hat{\tilde{f}}(\omega)}{\kappa_{i}+2\kappa-i\omega}.
\end{eqnarray}
Under the condition $\kappa_{i}=2\kappa$, if we only consider the normally ordered terms in the spectra of the $\hat{x}$ quadratures of the output fields, we find the results for the power spectra of the output fields and the squeezing spectra
\begin{eqnarray}
& &S_{aout}(\omega)=S_{bout}(\omega)=\frac{8\kappa^2+\omega^2}{16\kappa^2+\omega^2}N\frac{\Gamma^2}{\Gamma^2+\omega^2},\label{27in}\\
& &:S_{aout}^{x}(\omega):=:S_{bout}^{x}(\omega):=:S_{out}^{x}(\omega):=\frac{-8\kappa^2M\cos{\varphi}+(8\kappa^2+\omega^2)N}{16\kappa^2+\omega^2}\frac{\Gamma^2}{\Gamma^2+\omega^2}.\label{27}
\end{eqnarray}
We first of all observe that the output intensities are nonzero and hence no CPA occurs if one only confines to measurements of intensities. However there is a very interesting consequence of the interference if one looks at the measurements of the quadrature spectra. We discuss this now. For the special case $\varphi=0$, in the limit of large $r_{s}$, when $\omega=0$, Eq.~(\ref{27}) reduces to
\begin{eqnarray}\label{28}
:S_{out}^{x}(0):&=&0.5(N-M)=0.5(N-\sqrt{N(N+1)})\nonumber\\
&\approx&0.5[N-N(1+\frac{1}{2N})]=-0.25.
\end{eqnarray}
The normalized spectrum $:S_{out}^{x}(\omega):\times\frac{\Gamma^2+\omega^2}{\Gamma^2}$ of the output fields as a function of the normalized frequency $\omega/\kappa$ for different values of the squeezing parameter $r_s$ is shown in Fig.~\ref{Fig6}. One can see that $:S_{out}^{x}(\omega):\times\frac{\Gamma^2+\omega^2}{\Gamma^2}$ can be negative, so the squeezing happens in the $\hat{x}$ quadrature of the output fields, which is the signature of the quantum destructive interference between the quantum fields $\hat{a}_{in}(\omega)$ and $\hat{b}_{in}(\omega)$. Note that $:S_{out}^{x}(\omega):\times\frac{\Gamma^2+\omega^2}{\Gamma^2}$ has a dip at $\omega=0$, the depth of the dip is increasing with increasing the squeezing parameter $r_s$. For $r_{s}=3$ (solid curve), $:S_{out}^{x}(0):\approx-0.25$, which is in agreement with Eq.~(\ref{28}). Thus under the critical condition the output exhibits 50\% squeezing, this is a new signature of coherent perfect absorption when entangled fields are used. Note that in the absence of the resonator, $\hat{\tilde{a}}_{out}(\omega)=\hat{\tilde{a}}_{in}(\omega)$ and hence the output field shows no squeezing. Similar results can be obtained for the system of Fig.~\ref{Fig1}.

\section{Conclusions}
In conclusion, we have shown how to realize the CPA of single photons in a Fabry-Perot interferometer with a lossy medium and a micro-ring resonator with an internal loss. The key is to use the path entangled single photons. This can be achieved by using a single photon on a beam splitter. We also demonstrate that the entangled fields from a two-mode squeezed vacuum field can produce an analog of the CPA in the squeezing characteristics of the output fields. Our work shows how one can realize the filters for quantum fields.

\section*{Acknowledgments}
This work was supported by the Singapore National Research Foundation under NRF Grant No. NRF-NRFF2011-07.

\end{document}